\documentclass[11pt]{article}
\usepackage{hyperref}
\usepackage[blocks]{authblk}

\pdfoutput=1
\begin{document}
\title{Nonlinear spin-up of a thermally stratified fluid in cylindrical geometries}

\author[1,2]{J. Rafael Pacheco}
\author[2]{Sergey A. Smirnov}
\author[3]{Roberto Verzicco}

\affil[1]{Arizona State University, USA}
%%Environmental Fluid Dynamics, Department of Civil Engineering and 
%%Geological Sciences, The University of Notre Dame, South Bend, IN 46556. 
\affil[2]{The University of Notre Dame, USA}
\affil[3]{Texas Tech. University, USA} 
\affil[4]{Universita' di Roma ``Tor Vergata'', Italy}
        
%\\\vspace{6pt} Sibley School of Mechanical and Aerospace
%Engineering, \\ Cornell University, Ithaca, NY 14853, USA}
\maketitle
%% The abstract (in this file, and that submitted as text to arXiv) should
%% include the exact phrase
%% "fluid dynamics video" or "fluid dynamics videos"
\begin{abstract}
This is an entry for the Gallery of Fluid Motion of the 62nd Annual Meeting of the APS-DFD (fluid dynamics videos). 
This video shows the three-dimensional time-de\-pen\-dent 
incremental spin-up of a thermally stratified fluid in 
a cylinder and in an annulus.
The rigid bottom/side wall(s) are non-slip, and the upper surface is stress-free.
All the surfaces are thermally insulated.
The working fluid is water characterized by the kinematic viscosity $\nu$ and thermal
diffusivity $\kappa$.
Initially, the fluid temperature varies linearly with height and is characterized by a constant buoyancy frequency $N$,
which is proportional to the density gradient.
The system undergoes an abrupt change in the rotation rate from its initial value
$\Omega_i $, when the fluid is in a solid-body rotation state, to the final value $\Omega_f$.
Our study reveals a feasibility for transition from an axisymmetric initial circulation to
non-axisymmetric flow patterns at late spin-up times.
\end{abstract}
% main text
\section{Introduction}
This fluid dynamics video shows the 
baroclinic instability of the flow evolving in time due to an abrupt change of the
rotation rate of the cylinder/annulus from different perspectives in space. 
Our simulations revealed that azimuthal asymmetry is manifested in the
form of cyclonic and anticyclonic columnar eddies, which develop at the
temperature front formed by the highly distorted isotherms near the cylinder
sidewalls.
Strong deformation of the initial temperature field is caused by the
Ekman transport near the bottom. The front steepens until it reaches a
quasi-equilibrium state. The eddies grow in size and march along the
outer wall until they occupy a large portion of the tank. 
The characteristic Ekman number $E$ is small
and the Ekman bottom boundary layer is in the transitional state.
The Reynolds number based on the Ekman layer depth is $ Re_\delta = 90$  
and the Rossby number $\epsilon = (\Omega_f - \Omega_i) / \Omega_f$ (=0.727), 
so that the flow regime under investigation is highly nonlinear.
The details of the numerical method and validation tests are described in
\cite{SPV09} and references therein.
It is important to note that the formation of these large-scale eddies has been observed 
experiments of stratified fluids by salt \cite{SBBVS05,SBB05,Kan04}.
The variables shown in the video
are the temperature contours and isosurfaces of the 
$Q$-criterion \cite{HWM88} colored by temperature. 
The first section of the video presents the three-dimensional
numerical simulations for the cylinder, and the second section of the
video corresponds to the spin-up in the annulus.

The video animation file can be found at
\href{http://hdl.handle.net/1813/14148}{spinup.mpg}

\bibliographystyle{plain}

\end{document}